\documentclass[a4paper,11pt]{article}

\usepackage{aas_macros}
\usepackage{physics}
\usepackage[margin=2.5cm]{geometry}
\usepackage{graphicx}
\usepackage{subcaption}
\usepackage{amsmath,amssymb}
\usepackage{hyperref}
\usepackage{multirow}
\usepackage[dvipsnames,x11names,table]{xcolor}
\usepackage{rotating}

\hypersetup{
    colorlinks=true,
    urlcolor=orange,
    linkcolor=blue,
    citecolor=magenta}

\usepackage{sectsty}
\allsectionsfont{\sffamily}

\title{\textbf{\textsf{Cosmographic constraints on a Gödel-type rotating universe}}}

\author{Anshul Verma$^1$, Pavan K. Aluri$^1$, David F. Mota$^2$, Yuri N. Obukhov$^3$}
\date{\today}

\begin{document}

\maketitle
\centerline{$^1$Department of Physics, Indian Institute of Technology (BHU), Varanasi - 221005, India}
\centerline{$^2$Institute of Theoretical Astrophysics, University of Oslo, Sem Sælands vei 13, 0371 Oslo, Norway}
\centerline{$^3$Russian Academy of Sciences, Nuclear Safety Institute (IBRAE), B. Tulskaya str. 52, 115191 Moscow, Russia}

\begin{abstract}
We investigate the possibility of global cosmic rotation using a G\"odel-type rotating cosmological model, constrained through a cosmographic analysis of Type Ia supernovae (SNIa) from the Pantheon+ dataset. Employing a Taylor-expanded apparent magnitude--redshift relation derived via the Kristian-Sachs formalism, we analyze low-redshift SNIa data across five redshift bins (up to $Z \leq 0.5$). Our results reveal a mild but consistent preference for cosmic rotation, with the dimensionless rotation parameter $\Omega_0$ peaking at $0.29^{+0.21}_{-0.15}$ for $Z \leq 0.2$, and a broadly aligned anisotropy axis centered around equatorial coordinates $(243^\circ, -49^\circ)$. The inferred Hubble constant $h_0 \approx 0.73$ remains stable across all bins, while the deceleration parameter $q_0$ trends from near-zero to mildly negative values with increasing redshift. Model comparison using the Akaike Information Criterion (AIC) indicates a statistically significant preference for the rotating model over the standard $\Lambda$CDM cosmology at intermediate redshifts. These findings suggest that cosmic rotation, if present, may influence the late-time expansion history of the universe and warrants further investigation beyond the cosmographic regime.
\end{abstract}


\section{Introduction}
Rotation is an ubiquitous phenomena in any astronomical setting. Almost all heavenly bodies have some degree of rotation/spin, from the slow rotation of asteroids to the relativistic speeds at the center of galaxies. It is then natural to ponder over the question of whether the universe itself is rotating or not. Could the universe, on the largest scales, be rotating? Or whether the widespread phenomena of rotation is confined to regions where the universe is inhomogeneous? The question of cosmic rotation and its potential role in the structure and evolution of the universe has intrigued physicists for nearly a century.
Whittaker~\cite{1945whittaker} and Lemaître~\cite{1948lemaitre_atom_hypothesis} contemplated on the nature of a rotating ``primeval atom'' from which the universe supposedly emerged in relation to the formation of galaxies and large-scale structures.

The study of rotational effects in cosmology started with the work of Lanczos, who modeled the universe as a rigidly rotating dust cylinder with infinite radius~\cite{1924lanczos}, and later in the works of George Gamow~\cite{1946gamow}. However, a more realistic metric with rotation as a solution to Einstein's field equations was proposed by Gödel in 1949~\cite{1949godel}, that is given by the line element,
\begin{equation}
    ds^2 = a^2 \left( dt^2 - 2 e^x dtdy + e^{2x} dy^2 - dx^2 - dz^2 \right)\,,
    \label{eq:godel_metric_1949}
\end{equation}
where `$a$' is a positive number.
Gödel’s universe assumes dust as matter with an energy density $\varepsilon$ and a negative cosmological constant `$\Lambda$'. In this model, the angular velocity parameter `$\omega$' that quantifies cosmic rotation is given by $\omega^2 = 1/2a^2 = 4\pi G \varepsilon = - \Lambda$. This model provided a theoretical framework for studying rotating cosmologies for many years. Earlier model of Lanczos' was less physical compared to the Gödel's model~\cite{1936stockum}.

The metric in Eq.~(\ref{eq:godel_metric_1949}) is stationary and incorporates non-trivial global properties of rotating space-times. However, one fundamental problem with this solution was the existence of closed time-like curves (CTC) proved by Gödel himself~\cite{1952godel}. Gödel attempted to derive completely causal rotating cosmologies in his last work on rotation and presented a generalization of his original stationary model incorporating expansion, but without proof. First explicit solutions were reported in Refs.~\cite{1966maitra,1969ozsvath}. Maitra~\cite{1966maitra} extended the van Stockum's solution~\cite{1936stockum} with rotation as well as shear, and also introduced a straightforward condition for determining the absence of closed time-like curves, ensuring that the model avoids causality violations. Later, multiple solutions of rotating models with or without expansion have been developed, each offering unique insights into the role of rotation in cosmology. 
Theoretical advancements have given rotating cosmologies some serious considerations, including Einstein-Cartan theory that incorporates spin and torsion. Exact rotating solutions in Einstein-Cartan theory of gravity with spin and torsion were described, for example, in Refs.~\cite{1985bedran_cosmology_torsion,1985smalley_godel_cosmology,1986smalley_self_consistent_godel_cosmology,1986duarte_homogeneous_cosmos,1987obukhov_weyssenhoff_fluid,1988pavelkin_rotation_cosmology,1992korotky_obukhov_ecsk,1995palle_imperfect_fluid,1996palle_cosmological_observables,1998aman_riemann_cartan_space_times}.

The discovery of the cosmic microwave background (CMB) and its high degree of isotropy provided a very strong case for isotropic models. However, deviations from a near perfect isotropy are indicative of possible anisotropies present in the early stages of the evolution of our universe. This motivated a class of homogeneous but anisotropic cosmologies known as Bianchi type cosmological models~\cite{maccallum1969,maccallum1970,maccallum1971}. Gödel's solution is one such solution within the class of Bianchi type-III cosmological models. Homogeneous but anisotropic rotating cosmologies were verily analyzed earlier in the literature, in which upper limits on the value of cosmic rotation were obtained from data~\cite{1983barrow_nature,1985barrow_mnras,1969hawking_mnras,1973collins_hawking,1997kogut_prd,1970wolfe_apj}.
But, those studies did not separate the effects of vorticity and shear. In Refs.~\cite{1990obukhov,1992obukhov}, the effects of pure cosmic rotation were delineated from cosmic shear.

Even with the plethora of astrophysical data sets that were accumulated over the years, no completely convincing results were reported.
Birch~\cite{birch1982,1982birch_reply,1983conway_mnras} observed an apparent anisotropy in the alignment between the polarization vectors and the major axes of radio sources. He interpreted this observed anisotropy as potential evidence for cosmic rotation.
With the advent of precision era in cosmology, primarily driven by CMB observations, the Cosmological principle which is a foundational assumption of modern cosmology was put to test. This lead to finding many instances of isotropy violation in various astronomical and cosmological data sets (see for example Ref.~\cite{aluri2023cp}). Such instances of isotropy violation seen in CMB data have come to be known as \emph{CMB anomalies}~\cite{wmap7yranom,wmap9yrmaps,plk2013isostat,plk2015isostat,plk2018isostat}. There were many coincidental alignments among various preferred axes seen in these diverse data sets~\cite{johnjain2004}. Besides isotropy violations, the standard concordance model is also riddled with many \emph{tensions}~\cite{bull2016,abdalla2022,perivolaropoulos2022} from observations that challenge it despite its broad success. Apart from providing possible explanation to various signals of anisotropy seen in data, rotating models may also offer mild corrections to the Hubble flow, potentially providing an alternative explanation for dark energy and a possible resolution to the Hubble tension problem (for a review of Hubble tension and various possible (re)solutions see, for example, Ref.~\cite{Efstathiou:2020,valentino:2021,freedman2021,2023Kamionkowski}).

A study of spins in the intergalactic medium suggests that longest substantially rotating objects in the Universe are likely filaments of matter connecting neighboring galaxy pairs~\cite{2021Xia_MNRAS}. Another study reported observational evidence that cosmic filaments spin~\cite{2022Alexander}.
It was also reported that there is an asymmetry in the distribution of galaxy spin directions (handness)~\cite{Shamir2012,Shamir2020,Shamir2023}.
Further, some spectroscopic surveys indicate that the direction of rotation of a galaxy is coherent with the average motion of its neighbouring galaxies over a scale of 1~Mpc (megaparsec)~\cite{2019Lee,Lee2019}.
It was argued earlier that primordial turbulence or cosmic vorticity could have played a significant role in structure formation, providing a natural origin for galactic angular momentum~\cite{1972ApJ_Silk,1996Chernin}.
The availability of high-resolution CMB data in the last two decades saw a renewed interest in homogeneous but anisotropic cosmologies, including rotating models~\cite{2009Pontzen,coles2011}. Consequently, more stringent limits on cosmic shear and vorticity were derived using CMB~\cite{Jaffe2005ApJ,jaffe2006,saadeh2016MNRAS}.


In Sec.~\ref{sec:godel-model}, we layout the general properties of a rotating and expanding form of Gödel metric given in Eq.~(\ref{eq:godel_metric_1949}) and present the relation between apparent magnitude of a luminous source `$m$' vs redshift `$Z$' in such a model. In Sec.~\ref{sec:data-likelihood-fn}, the data used in constraining the Gödel-like model under study and the methodology to derive constraints on its parameters are described. Finally, we present our results in Sec.~\ref{sec:Results} and summarize our finding in Sec.~\ref{sec:conclusion}.

\section{Gödel-type rotating cosmological model}
\label{sec:godel-model}
In this section we briefly discuss the Gödel-like model that we are going to investigate in this work. First we present an overview of Gödel-type metric with additional details as necessary, followed by a description of the Taylor expanded apparent magnitude vs redshift relation in such a universe. For more details the reader may consult some of the original works and the references therein~\cite{1990obukhov,1991KorotkyObukhov,1996obukhov,2000obukhov}.

\subsection{Gödel-type metric}
\label{ssec:godel-like-metric}

A wide class of homogeneous, sheer-free cosmological models with expansion and rotation were studied in Refs.~\cite{1990obukhov,1991KorotkyObukhov}. These space-time manifolds are completely causal. The cosmic  microwave background (CMB) radiation remains isotropic, and rotation does not produce additional astrophysical parallax effects. The general non-stationary form of the original Gödel metric given by Eq.~(\ref{eq:godel_metric_1949}) is described by the invariant space-time interval,
\begin{equation}
    ds^2 = dt^2 - 2\sqrt{\sigma} R(t) e^{mx} \, dt \, dy - R^2(t) \left( dx^2 + k e^{2mx} dy^2 + dz^2 \right),
    \label{eq:go_metric}
\end{equation}
where $R(t)$ (or simply `$R$') is the time dependent scale factor (same as $a(t)$ in FLRW metric) and $m, \sigma, k > 0$ are constants. We employed natural units in the above equation so that the speed of light is set to unity ($c=1$). The condition $k > 0$ ensures the absence of closed time-like curves in this model. Therefore the metric in Eq.~(\ref{eq:go_metric}) represents a Gödel-type model that features both rotation and expansion. Since the determinant of the metric above is $| g_{\mu \nu} |= -(k + \sigma)e^{2mx}$, in terms of geometry, one requires $(k+\sigma) > 0$ for it to conform with the Lorentzian signature, that is trivially satisfied by the positivity condition on the constants $m, \sigma$, and $k$ as mentioned above.

The magnitude of the global rotation along the $z$-axis (the axis of anisotropy) can be found to be given by,
\begin{equation}
\omega = \frac{m}{2R} \sqrt{\frac{\sigma}{k + \sigma}}\,.
\label{eq:godel-rot-par}
\end{equation}
It is clear that the rotation parameter  $\omega$ decreases as the universe expands.

In order to make theoretical predictions and compare them with the observations, one needs to understand how electromagnetic radiation propagates through space-times following the null geodesic path. The geodesic equations as usual follow:
\begin{equation}
    k^\mu \nabla_\mu k^\nu = 0, \quad k^\mu k_\mu = 0,
    \label{eq:geodesic_eqn}
\end{equation}
where $k^\mu = {dx^\mu}/{d\lambda}$ is the tangent vector to the geodesic curve $x^\mu(\lambda)$ parameterized by an affine parameter `$\lambda$'.  These equations describe how light paths evolve in a rotating cosmological background and provide the mathematical basis for understanding observations in such universes.

Next, we describe the luminosity distance (or equivalently the distance modulus) versus the redshift relation that is required to test this model against observations.

\subsection{Apparent magnitude versus redshift $(m-Z)$ relation}
\label{ssec:m_vs_z_relation}

Owing to spatial homogeneity of this model, we can assign the location of the observer as $P = (t_0,0,0,0)$ without loss of generality. Now, the apparent position of a source of radiation as seen by an observer at `$P$' can be denoted by the usual spherical-polar coordinates $(\theta,\phi)$ on the celestial sky.

The observable effects of a rotating model on cosmological scales include the determination of the Hubble parameter $H_0 = (\dot{R}/R)_P$, the rotational parameter $\omega_0 = \omega(t_0)$, and the deceleration parameter $q_0 = -(\ddot{R}R/\Dot{R}^2)_P$. This section will present the potential observational consequences of cosmic rotation within the framework of the Gödel-type universe of Eq.~(\ref{eq:go_metric}).

Observational tests, like those relating apparent magnitude and redshift $(m - Z)$, angular size with redshift, and other such relations, allows one to probe  how astrophysical observables are influenced by a source's location $(\theta, \phi)$ in the sky with respect to the axis of anisotropy in a rotating cosmological model. To perform cosmological tests with observations from a radiation source such as Type Ia supernova, an important test is comparing the theoretical area distance $r$, which measures the intrinsic area $dA_S$ of a source `$S$' subtending a solid angle $d\Omega_P$ at the observer's location `$P$'. The area distance `$r$' between `$P$' and `$S$' is given by,
\begin{equation}
dA_S = r^2 \, d\Omega_P\,.
\end{equation}
This distance depends both on the direction of observation $(\theta,\phi)$, and on `$\lambda$' the affine parameter i.e., $r=r(\lambda;\theta,\phi)$ (or equivalently, the emission time `$t$' of the source that is detected by the observer at time $t_0$ i.e., $r=r(t;\theta,\phi)$).

The relation between luminosity distance `$\mathcal{D}$' and the area distance `$r$' is $\mathcal{D}=r(1+Z)^2$. Further the distance modulus `$\mu$' is related to the luminosity distance via the apparent magnitude of an astronomical source as,
\begin{equation}
    \mu = m - M = 5\log_{10}\left( \frac{\mathcal{D}}{10~pc}\right)\,.
    \label{eq:dist-mod-defn}
\end{equation}
where `$M$' is the absolute magnitude of a radiating source at a fixed distance of $10~pc$ (parsec) from the source.

Rotational effects are most prominent near the rotation axis, where simple expressions can be derived. For arbitrary directions, the formulas become complex and are often simplified using the \emph{Kristian-Sachs formalism}~\cite{1966Kristian_Sachs_ApJ}. This formalism allows one to express geometric and physical quantities of interest as a power series in terms of affine parameter or redshift, enabling one to derive the magnitude-redshift relation. Specifically, we can get a power series expansion for `$Z$' in terms of `$r$' as
\begin{equation}
1 + Z= \sum_{n=0}^\infty a_n r^n\,,
\end{equation}
where `$a_n$' are coefficients of this Taylor expansion such that $a_0=1$ for consistency when $Z=0$. This can be inverted to find a relation between `$r$' and `$Z$' as,
\begin{equation}
    r = \sum_{m=1}^\infty b_m(\theta,\phi,\boldsymbol{\xi}) Z^m\,,
\end{equation}
where the coefficients of expansion `$b_m$' (as well as $a_n$) are functions of $(\theta,\phi)$ and the cosmological parameters $\boldsymbol{\xi}=H_0$, $q_0$, $\omega_0$, etc. Therefore, these expansions capture the angular dependence of the effects of rotation on cosmological observables.

Using this inverted relation for area distance, $r=r(Z;\theta,\phi,\boldsymbol{\xi})$, in Eq.~(\ref{eq:dist-mod-defn}) along with $\mathcal{D}=r(1+Z)^2$, the apparent magnitude-redshift relation ``$m-Z$'' in the Gödel-type rotating universe of Eq.~(\ref{eq:go_metric}) can be found to be given by~\cite{1996obukhov,2000obukhov},
\begin{align}
    \mu^{\rm th} = m - M &=  \mu_0-5\log_{10}{h_0} + 5\log_{10}Z + \frac{5}{2}(\log_{10}e)(1 - q_0)Z \nonumber \\
    &\quad - 5\log_{10}{\left( 1 + \sqrt{\frac{\sigma}{k + \sigma}}\sin{\theta}\sin{\phi} \right)} \nonumber \\
    &\quad - \frac{5}{2}(\log_{10}e)\frac{\Omega_0}{h_0} \,\, \frac{\sin{\theta}\cos{\phi} \left( \sqrt{\frac{\sigma}{k + \sigma}} + \sin{\theta}\sin{\phi} \right)}{\left(1 + \sqrt{\frac{\sigma}{k+\sigma}}\sin{\theta}\sin{\phi} \right)^2}Z + O(Z^2)\,,
\label{eq:godel_distance_modulus}
\end{align}
up to linear order in `$Z$'.
Here `$h_0$' is dimensionless Hubble parameter at current time in units of $u=100$~km/s/Mpc i.e., $H_0 =u\,h_0$. The parameter $\Omega_0$ is also dimensionless and quantifies the magnitude of global rotation today. The constant offset $\mu_0$ arises as a result of all dimensionful parameters in the luminosity distance vs redshift relation viz., $\mu_0 = 5\log_{10}(c/(u\times10~pc))\approx42.384$, and `$c$' is the speed of light in km/s. Hereafter we use $\tilde{\rho}=\sqrt{\sigma/(k+\sigma)}$ for brevity.

This expression shows how cosmic rotation could effect the magnitude-redshift relation as compared to the standard FLRW case. We will use this theoretical distance modulus $\mu^{\rm th}$ to compare with apparent magnitudes of Type Ia supernovae from observations to derive parameter constraints by performing an MCMC analysis as described in the next section. 

\section{Type Ia supernova data and Parameter estimation methodology}
\label{sec:data-likelihood-fn}

The Pantheon+ Type Ia supernova (SNIa) dataset\footnote{\url{https://github.com/PantheonPlusSH0ES/DataRelease}} contains 1701 light curves corresponding to approximately 1550 unique spectroscopically confirmed SNIa~\cite{scolnic2022}. The remaining light curves in the dataset pertain to either the same supernova (SN) observed by different surveys or to ``SN siblings'' that are SNIa found within the same host galaxy. This latest data set includes significant enhancements, such as the addition of new SNIa and corrections for peculiar velocity effects and host galaxy bias in the SNIa covariance matrix.

In order to measure the luminosity distance to these supernovae (SNe), we use Cepheid variable stars, a type of pulsating stars whose intrinsic luminosity is related to their period of variability. This relationship, known as the period-luminosity (P-L) relation, was first discovered by Henrietta Swan Leavitt in 1912~\cite{Leavitt:1908vb,Leavitt1912} while studying stars in the Magellanic Clouds. Cepheids are bright enough to be observed in relatively nearby galaxies. Distances to them within the Milky Way were first measured using parallax (a geometric method). This provides a way to estimate the absolute brightness (or luminosity) of these stars based solely on their observed period. Next, Cepheids in nearby galaxies - those that also host Type Ia supernovae (SNe Ia) - are observed to link the supernova luminosity scale to absolute distances. This step is crucial because once the absolute magnitudes of SNe are calibrated using Cepheid distances, SNe Ia can then be used to measure much greater cosmological distances. This calibration process allows us to determine the luminosity distance to very distant supernovae in the Hubble flow (of the expanding universe). The Pantheon+ dataset contains Cepheid anchors that were part of the Pantheon+SH0ES analysis~\cite{Riess2022SH0ES}. 42 SNe Ia from Cepheid-calibrated host galaxies are included in the analysis, along with 277 Hubble-flow supernovae.

\begin{figure}
    \centering
    \includegraphics[width=0.34\textwidth]{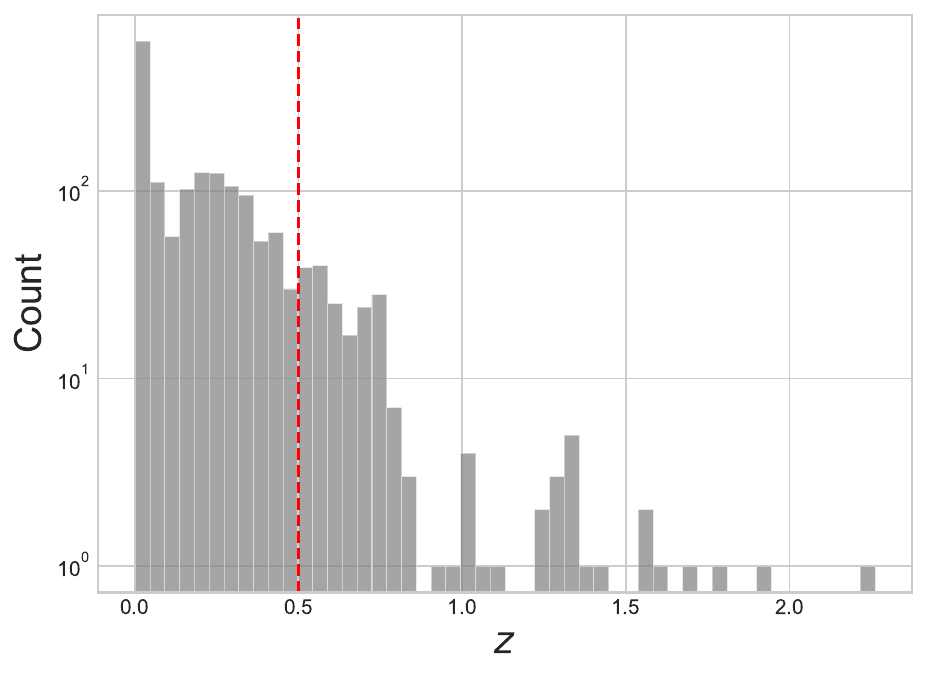}
    ~
    \includegraphics[width=0.63\textwidth]{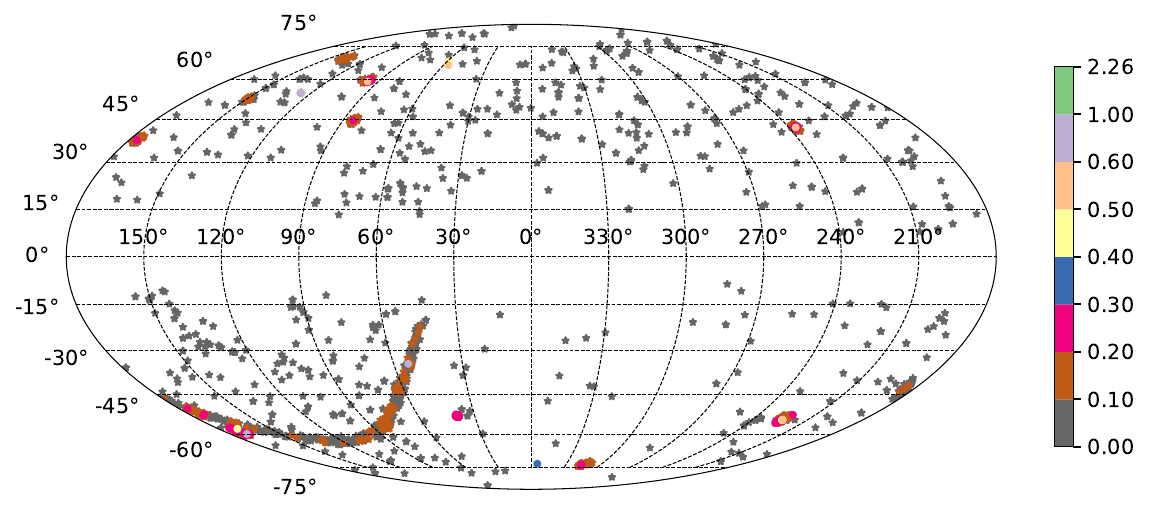}
    \caption{\emph{Left :} Number count histogram of SNIa in Pantheon+ data with respect to redshift in 50 bins. The vertical dashed line indicated the maximum redshift $Z\leq0.5$ of SNe Ia objects considered in our cosmographic study. \emph{Right :} Sky distribution of Pantheon+ SNIe in galactic coordinates. SNe Ia with $Z\leq0.5$ are denoted by a $\star$ and those beyond that redshift are denoted by $\circ$.}
    \label{fig:pplus_dist}
\end{figure}

SNIa cosmology relies on comparing the theoretical relationship between redshift ($Z^*$) and luminosity distance ($\mathcal{D}$) or equivalently the distance modulus ($\mu^{\rm th}$) that is dependent on various cosmological parameters of our model, with the observed distance modulus ($\mu^{\rm obs}$) to derive constraints on them. The theoretical and observed distance modulus can be formally expressed as
\begin{equation}
    \mu^{\rm th}({\boldsymbol\xi},Z^*) = 5\log_{10}\left( \frac{\mathcal{D}({\boldsymbol\xi},Z^*)}{10~pc}\right)\,, \quad \mu^{\rm obs} = m_{B,\rm corr} - M_0\,,
    \label{eq:distmod}
\end{equation}
where `$Z^*$' is the redshift of an SNIa, ${\boldsymbol\xi}=\{h_0, \omega_0, q_0,\tilde{\rho}, (R_a,D_a)\}$ are the model parameters to be constrained, and `$pc$' stands for the distance unit `parsec'. Here $(R_a,D_a)$ is the direction of anisotropy i.e., axis of cosmic rotation in equatorial coordinates.
Further $m_{B,\rm corr}=m_B+\alpha x_1-\beta \texttt{c} + \gamma G_{\rm host}$ is the corrected apparent $B$-band magnitude of an SNIa from observations for its correlation with light curve fitting parameters viz., stretch ($x_1$), color (\texttt{c}) and host mass bias ($G_{\rm host}$). The corresponding coefficients $\alpha$, $\beta$ and $\gamma$ are global fit parameters along with the absolute $B$-band magnitude `$M_0$' that are conventionally assumed to be same for all SNIa. `$M_0$' is notably degenerate with the Hubble parameter $H_0$ (i.e., $h_0$), underscoring the importance of using SH0ES Cepheid variables in breaking that degeneracy.

The $\chi^2({\boldsymbol \Xi})$ 
 function, which depends on various fitting parameters ${\boldsymbol \Xi}$, is defined as,
\begin{equation}
\chi^2({\boldsymbol \Xi}) = \Delta{\boldsymbol \mu}^T ({{\bf C}_{\rm stat+sys}^{\rm SN} + {\bf C}_{\rm stat+sys}^{\rm Cepheid}})^{-1} \Delta{\boldsymbol \mu}\,,
\label{eq:chi2-cov}
\end{equation}
where the goal is to minimize this function to derive the desired constraints.
Here, $\Delta{\boldsymbol \mu}=(\Delta \mu_1, \Delta \mu_2, \ldots, \Delta \mu_n)^T$ is a column vector with `$n$' elements corresponding to total number of SNe Ia in the dataset. The residual for the $i$-th SNIa is given by,
\begin{equation}
    \Delta \mu_i =
    \left\{
\begin{array}{l l}
    \mu^{\rm obs}_i - \mu_i^{\rm Cepheid}, & \text{if } i \in \text{Cepheid hosts}\,, \\
    \mu_i^{\rm obs} - \mu^{\rm th}({\boldsymbol\xi},Z^*_i), & \text{otherwise}\,,
\end{array}
\right.
\label{eq:dist-mod-resd}
\end{equation}
where,
\begin{equation}
\mu^{\rm obs}_i=m_{B,\rm corr}^i - M_0=m^i_{B} + \alpha x_{1,i} - \beta \texttt{c}_i + \gamma G_{{\rm host},i} - M_0\,,    
\end{equation}
and $\mu^{\rm th}_i$ denotes the theoretical distance modulus dependent on the model parameters for that $i$-th SNIa, . The above relation (Eq.~\ref{eq:dist-mod-resd}) implies that whenever an SNIa host galaxy also contains a Cepheid, then the theoretical distance modulus $\mu^{\rm th}_i$ is replaced by the well measured value of distance modulus `$\mu^{\rm Cepheid}_i$' to that galaxy obtained using the tight Cepheid period-luminosity relation. This way the degeneracy between the (dimensionless) Hubble parameter `$h_0$' and the absolute magnitude `$M_0$' is removed.

Apart from the global fit parameters `$\alpha$' and `$\beta$' corresponding to stretch and color correction of a supernova light curve, another dimensionless parameter `$\gamma$' captures the dependence of apparent magnitude `$m_B$' on host-galaxy stellar mass viz., $G_{{\rm host},i}$. This host-galaxy stellar mass parameter is defined as a step function depending on its mass as \( G_{{\rm host},i} = +\frac{1}{2} \) if \( M_{{\rm host},i} > 10^{10} M_{\odot} \) and \( G_{{\rm host},i} = -\frac{1}{2} \) if \( M_{{\rm host},i} < 10^{10} M_{\odot} \), where $M_{\odot}$ denotes 1 solar mass. The values of $x_{1,i}$, $\texttt{c}_i$, $m_B^i$, $G_{{\rm host},i}$ (as $\log_{10}(M_{{\rm host},i}/M_{\odot})$) and $\mu_i^{\rm Cepheid}$ are provided as part of Pantheon+ dataset for each SNIa.

Note that Pantheon+ collaboration also provides $m_{B,\rm corr}^i$ for each SNIa. Instead, we use $m_B^i$ and derive constraints on $\alpha$, $\beta$ and $\gamma$ along with cosmological parameters of our model.
To account for the correlations among SNIa, which are influenced by both statistical and systematic uncertainties in the light curve fitting, we utilize the full covariance matrix.
${\bf C}_{\rm stat+sys}^{\rm SN}$ represents the covariance matrix for supernovae, while ${\bf C}_{\rm stat+sys}^{\rm Cepheid}$ denotes the covariance matrix for the SH0ES Cepheid-host distances, both of which are also provided as part of the Pantheon+ data release.

\section{Results}
\label{sec:Results}
This section presents our results from an MCMC likelihood analysis of the Gödel-type rotating cosmological model, discussed in Sec.~\ref{sec:godel-model}, to probe cosmic rotation.  The full set of parameters to be constrained are ${\boldsymbol\Xi}=\{{\boldsymbol\xi}, M_0 , \alpha, \beta, \gamma\} =\{h_0, \Omega_0, q_0,\tilde{\rho}, (R_a,D_a),M_0, \alpha, \beta, \gamma\}$ along with the cosmological parameters ${\boldsymbol\Xi}$ that describe the model of our present study. Since the luminosity-distance vs redshift relation given in Eq.~(\ref{eq:godel_distance_modulus}) is a Taylor expanded relation up to linear order in redshift, $Z$, we derive \emph{cosmographic} constraints on these parameters using low redshift SNIa data up to a maximum of $Z=0.5$. Specifically we estimate these parameters in five redshift bins ranging from $Z \leq 0.1$, $Z \leq 0.2$, $Z \leq 0.3$, $Z \leq 0.4$, and $Z \leq 0.5$. The astrophysical parameters $\{ M_0, \alpha, \beta, \gamma\}$ correspond to the SNIa absolute magnitude and coefficients of SNIa light-curve fitting parameters related to stretch, color, and host galaxy properties respectively. Actual cosmological parameters of our model are ${\boldsymbol\xi}=\{h_0, \Omega_0, q_0,\tilde{\rho}, (R_a,D_a)\}$. All these parameters are constrained using the latest Pantheon+ Type Ia supernova dataset.

\begin{figure}
\centering
\includegraphics[width=0.45\textwidth]{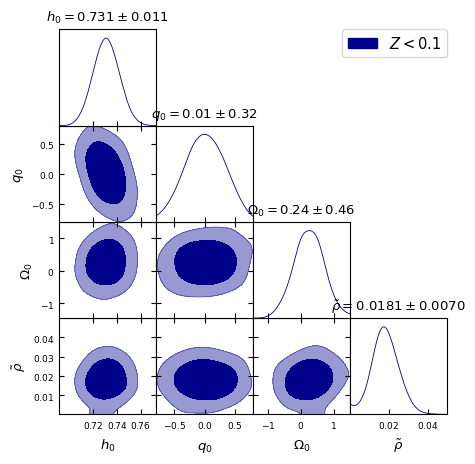}
~
\includegraphics[width=0.45\textwidth]{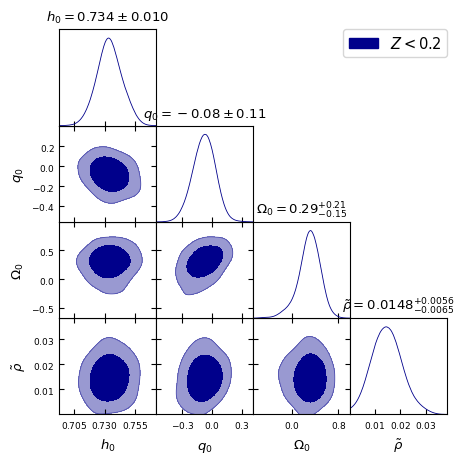}
~
\includegraphics[width=0.45\textwidth]{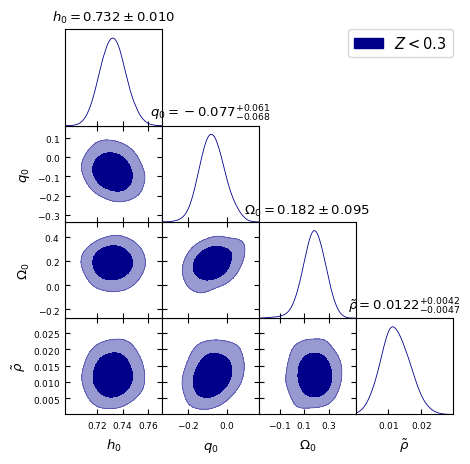}
~
\includegraphics[width=0.45\textwidth]{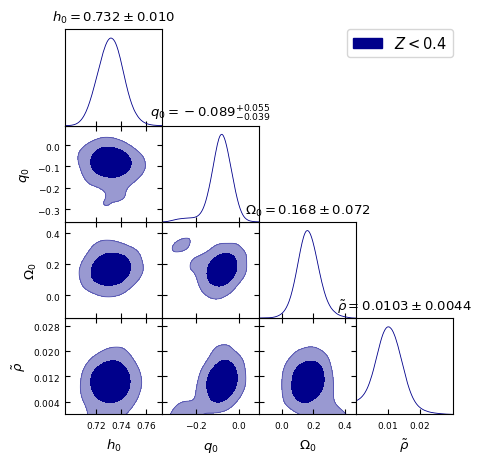}
~
\includegraphics[width=0.45\textwidth]{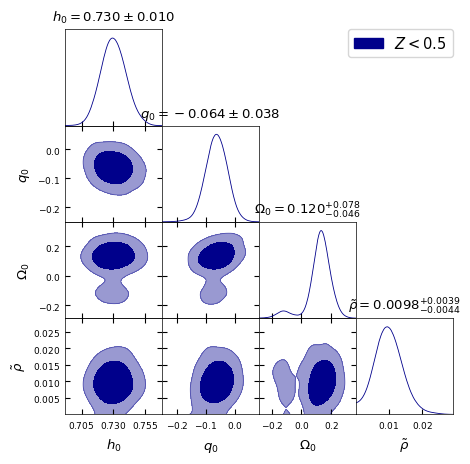}
~
\includegraphics[width=0.45\textwidth]{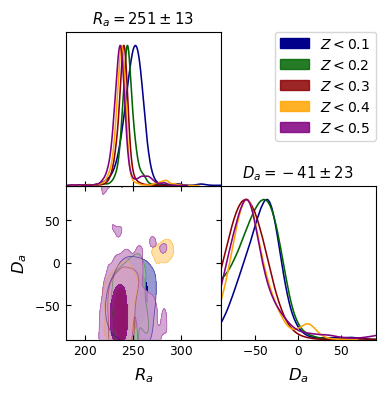}
\caption{2D contour plots of cosmological parameters $h_0$, rotation velocity parameter $\Omega_0$, deceleration parameter $q_0$, and $\tilde{\rho}$ also related to cosmic rotation corresponding to the Gödel-type model with expansion and rotation. All the parameters shown are dimensionless. \emph{Bottom right} plot is the density plot showing constrains on the inferred anisotropy axis of rotation for all redshift ranges.}
\label{fig:godel-constr-mcmc}
\end{figure}

\begin{figure}
\centering
\includegraphics[width=0.45\textwidth]{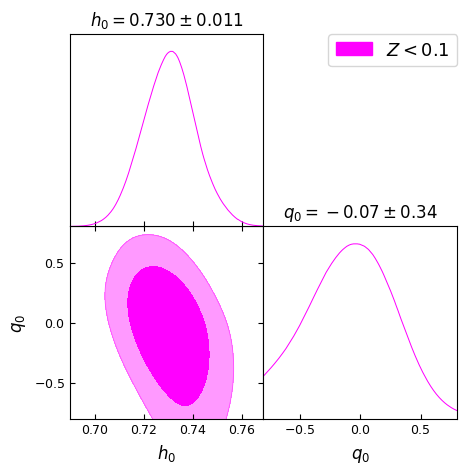}
~
\includegraphics[width=0.45\textwidth]{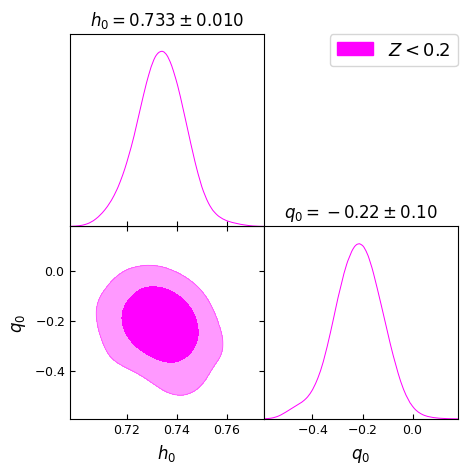}
~
\includegraphics[width=0.45\textwidth]{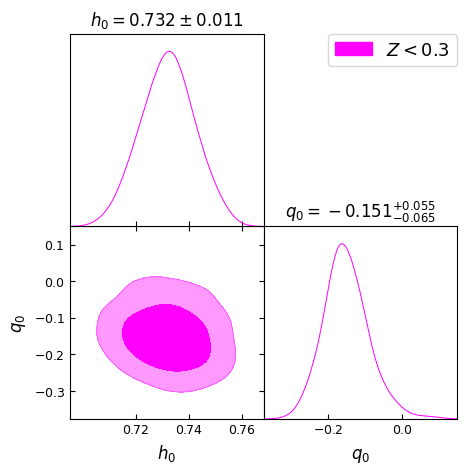}
~
\includegraphics[width=0.45\textwidth]{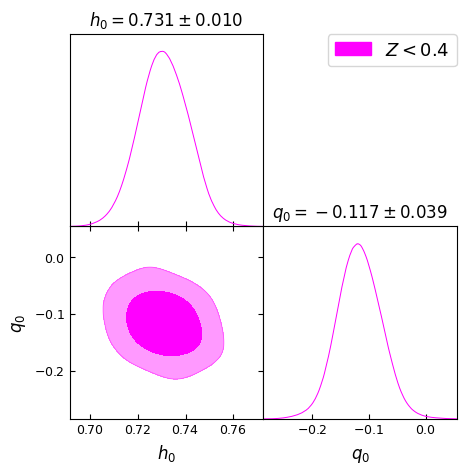}
~
\includegraphics[width=0.45\textwidth]{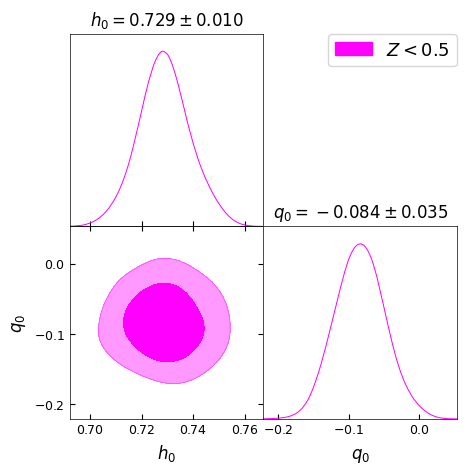}
~
\includegraphics[width=0.45\textwidth]{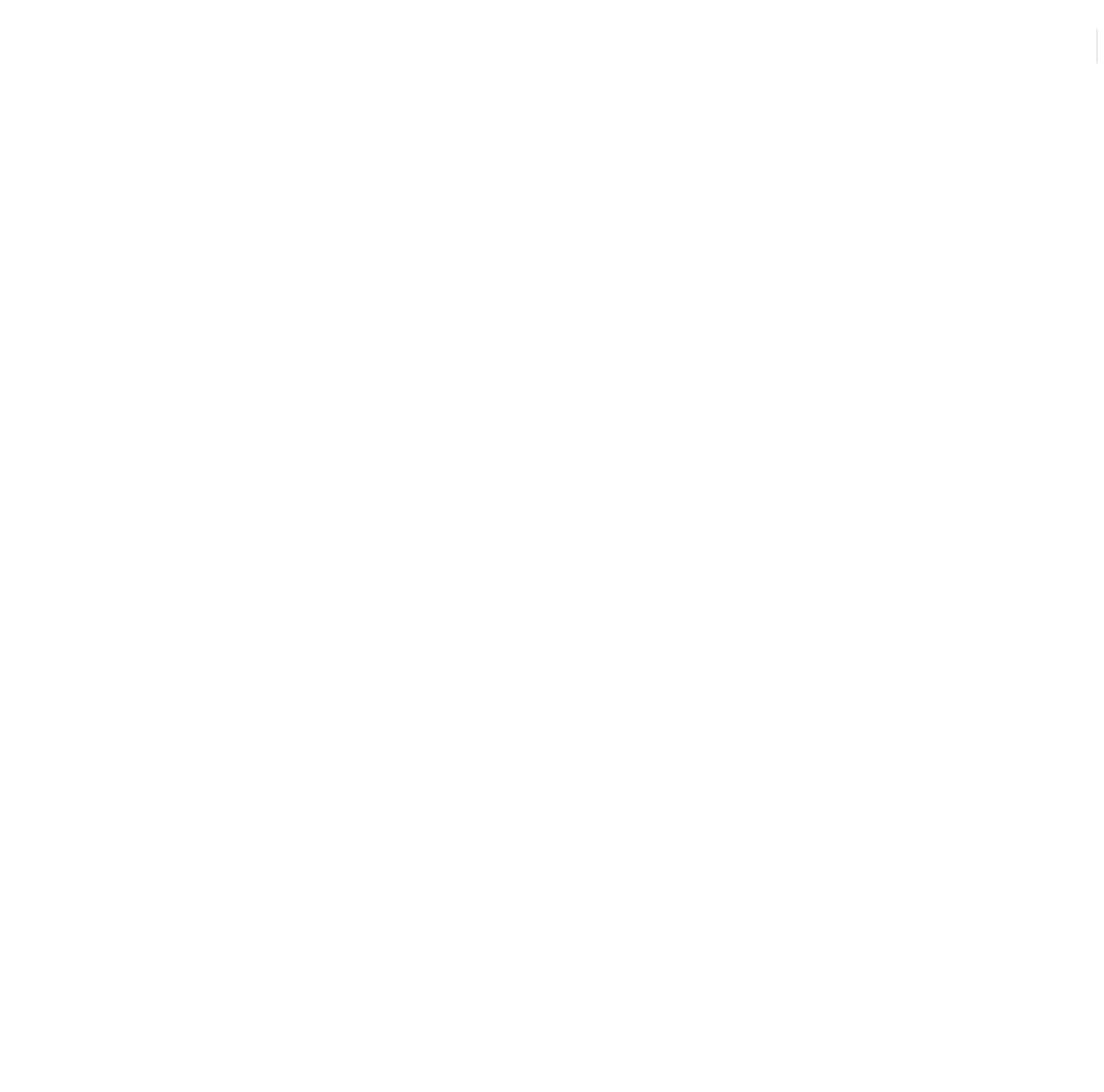}
\caption{2D contour plots of cosmological parameters $h_0$ and deceleration parameter $q_0$ corresponding to the standard $\Lambda$CDM model with only expansion.}
\label{fig:lcdm-constr-mcmc}
\end{figure}

In Fig.~[\ref{fig:godel-constr-mcmc}], we present the 2D contour plots illustrating the $1\sigma$ and $2\sigma$ confidence levels (CL) for the model parameters $h_0, \Omega_0, q_0$, and $\tilde{\rho}$ from different redshift bins mentioned above.
It is important to note that these values should not be directly compared with $\Lambda$CDM constraints from SH0ES and Pantheon+ collaborations~\cite{Riess2022SH0ES,Brout2022Pantheonplus}, as the latter were obtained using the full dataset containing SNe Ia upto $Z \lesssim 2.3$. Thus, we also perform a similar cosmographic analysis of standard $\Lambda$CDM model. These results are presented in Fig.~[\ref{fig:lcdm-constr-mcmc}] to make appropriate comparison by selected red-shift bins. All the parameter constraints derived from the two models for the chosen redshift bins are tabulated in Table~\ref{tab:cosmo-par-compare}.

The Hubble constant `$h_0$' in the G{\"o}del-type model as well as $\Lambda$CDM model remains essentially same across all five redshift bins, with $h_0 \approx 0.73$ for $Z \leq 0.1$, $\leq 0.2$, $\hdots$, $\leq 0.5$. These results agree well with the SH0ES determination of $H_0 = 73.04 \pm 1.04$ (km/s/Mpc)~\cite{Riess2022SH0ES}. The agreement suggests two possibilities, that the determination of `$h_0$' using Pantheon+  SNe Ia compilation is driven by the tighter constraining power of SH0ES subset (that helps breaking the degeneracy between $h_0$ and $M_0$) or that the introduction of rotation in the Gödel-type framework may not be affecting the (local) expansion rate significantly.

The maximum likelihood values of the deceleration parameter `$q_0$', shown in Fig.~[\ref{fig:godel-constr-mcmc}], for the Gödel-type model are $q_0 \approx 0.01$, $-0.08$, $-0.077$, $-0.089$, and $-0.064$ for the five redshift ranges $Z \leq 0.1$, $\hdots$, $\leq 0.5$, respectively. A small positive value for $q_0$ at $Z \leq 0.1$ may be due to the limited constraining power of SNe Ia at low redshift or the influence of mild rotational effects on $q_0$ or nature of local matter distribution. 
It is clear though, from Table~\ref{tab:cosmo-par-compare}, that $q_0$ is consistent with `zero' deceleration/acceleration when using SNe Ia upto $Z\leq0.2$, given the error bars.
However, as we include more number of supernovae with higher redshifts, it tends to take negative values consistent with $\Lambda$CDM predictions. Here we note that SH0ES team reported $q_0 = -0.510\pm 0.024$ that includes some intermediate redshift SNe also (up to $Z<0.8$) in their data from the Pantheon+ sample~\cite{Riess2022SH0ES}. In comparison, as shown in Fig.~[\ref{fig:lcdm-constr-mcmc}], the $\Lambda$CDM model with no rotation exhibits consistently negative values, inline with the standard model expectations. We find $q_0\approx-0.07$, $-0.22$, $-0.151$, $-0.117$, and $-0.084$ for the same redshift ranges, though $q_0$ is consistent with `zero' given the error bars in the first redshift bin. We observe that these estimates of $q_0$ from $\Lambda$CDM model are smaller (more negative) than the values found with Gödel-type model for our Universe. Thus, this may be indicative of the (local) effects of rotation vis-a-vis expansion.

\begin{sidewaystable}
    \centering
    \renewcommand{\arraystretch}{1.5}
    \begin{tabular}{|c|c|c|c|c|c|}
        \hline
        \multirow{2}{*}{\textbf{Parameter}} & \multicolumn{5}{|c|}{\textbf{Redshift bins (\#SNIa)}} \\ \cline{2-6}
        & $Z \leq 0.1 \ (\#741)$ & $Z \leq 0.2 \ (\#948)$ & $Z \leq 0.3 \ (\#1207)$ & $Z \leq 0.4 \ (\#1393)$ & $Z \leq 0.5 \ (\#1491)$ \\ 
        \hline
        \hline
        \rowcolor{white}
        \multicolumn{6}{|c|}{Gödel-type rotating model} \\
        \hline
        \rowcolor{AntiqueWhite1}
        $h_0$ & $0.731\pm 0.011            $ & $0.734\pm 0.010            $ & $0.732\pm 0.010            $ & $0.732\pm 0.010            $ & $0.730\pm 0.010            $ \\
        \hline
        $q_0$ & $0.01\pm 0.32              $ & $-0.08\pm 0.11             $ & $-0.077^{+0.061}_{-0.068}  $ & $-0.089^{+0.055}_{-0.039}  $ & $-0.064\pm 0.038           $ \\
        \hline
        \rowcolor{AntiqueWhite1}
        $\Omega_0$ & $0.24\pm 0.46              $ & $0.29^{+0.21}_{-0.15}      $ & $0.182\pm 0.095            $ & $0.168\pm 0.072            $ & $0.120^{+0.078}_{-0.046}   $ \\ 
        \hline
        $\Omega_0/h_0{}^*$ & $0.33\pm 0.63$ & $0.40^{+0.28}_{-0.21}$ & $0.25\pm 0.13$ & $0.230\pm 0.098$ & $0.16^{+0.11}_{-0.06}$ \\

        \hline
        \rowcolor{AntiqueWhite1}
        $\tilde{\rho}$ & $0.0181\pm 0.0070          $ & $0.0148^{+0.0056}_{-0.0065}$ & $0.0122^{+0.0042}_{-0.0047}$ & $0.0103\pm 0.0044          $ & $0.0098^{+0.0039}_{-0.0044}$ \\ 
        \hline
        {$(R_a, D_a)$} & $(251\pm 13, -41\pm 23)$ & $(244.8^{+5.7}_{-6.6}, -46\pm 22)$ & $(239.6^{+5.9}_{-4.7}, -56^{+15}_{-26})$ & $(239.3^{+3.8}_{-7.2}, -52^{+12}_{-28})$ & $(238.1^{+4.6}_{-8.8}, -52^{+10}_{-31})$ \\
        \hline
        \rowcolor{AntiqueWhite1}
        $\chi^2$ & $685.38$ & $893.58$ & $1139.64$ & $1304.09$ & $1407.49$ \\
        \hline
        \hline
        \rowcolor{white}
        \multicolumn{6}{|c|}{$\Lambda$CDM} \\
        \hline
        \rowcolor{AntiqueWhite1}
        $h_0$ & $0.730\pm0.011$ & $0.733\pm0.010$ & $0.732\pm0.011$ & $0.731\pm0.010$ & $0.729\pm0.010$ \\ 
        \hline
        $q_0$ & $-0.07 \pm 0.34$ & $-0.22\pm 0.10$ & $-0.151^{+0.055}_{-0.065}$ & $-0.117\pm0.039$ & $-0.084\pm0.035$ \\
        \hline
        \hline
        \rowcolor{AntiqueWhite1}
        $\chi^2$ & $693.54$ & $903.82$ & $1152.24$ & $1315.22$ & $1419.36$ \\
        \hline
        \hline
        \rowcolor{white}
        \multicolumn{6}{|c|}{Model comparison - AIC} \\ \hline
        \rowcolor{AntiqueWhite1}
        $\Delta \chi^2$ & $8.15$ & $10.23$ & $12.60$ & $11.12$ & $11.87$ \\
        \hline
        $\Delta\mathcal{A}$ & $-0.15$ & $-2.23$ & $-4.60$ & $-3.12$ & $-3.87$ \\
        \hline
        \rowcolor{AntiqueWhite1}
        $\mathcal{P}$ & $1.07$ & $3.06$ & $9.98$ & $4.77$ & $6.92$ \\ \hline
    \end{tabular}
    \caption{Comparison of cosmological parameters for different redshift ranges between the Gödel-type model and $\Lambda$CDM, including differences in $\chi^2$ i.e $\Delta\chi^2$, relative AIC i.e., $\Delta\mathcal{A}$ and how probable one model is over the other i.e., $\mathcal{P}$ values. The number of free parameters in the rotating model under study are $k=10$, while for the reference $\Lambda$CDM model is $k=6$.}
    \label{tab:cosmo-par-compare}
\end{sidewaystable}

\begin{figure}
    \centering
    \includegraphics[width=0.9\textwidth]{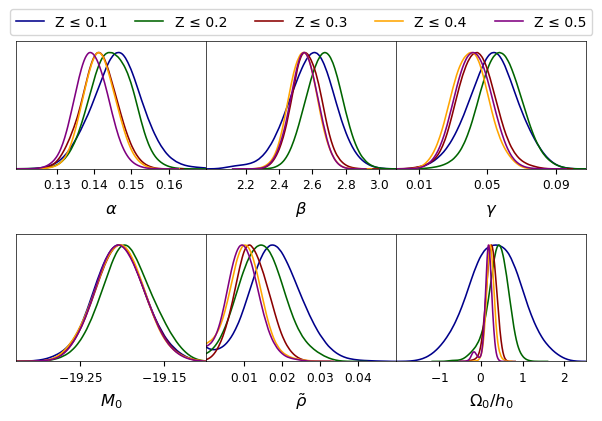}
    \caption{\emph{Top}: 1D posterior distribution of the nuisance parameters
    	$\alpha$, $\beta$ and $\gamma$ related to SNIa light curve fitting
    	that are fit along with the cosmological model parameters.
    \emph{Bottom}: The 1D posterior distribution of SNIa absolute magnitude
    `$M_0$' (\emph{left}), and the cosmological parameters `$\tilde{\rho}$' and
    the derived parameter $\Omega_0/h_0$ (\emph{middle} and \emph{right}) that
    are dimensionless related to global rotation.}
    \label{fig:1D_rho_M0_omega_by_h}
\end{figure}

\begin{figure}
    \centering
    \includegraphics[scale=0.7]{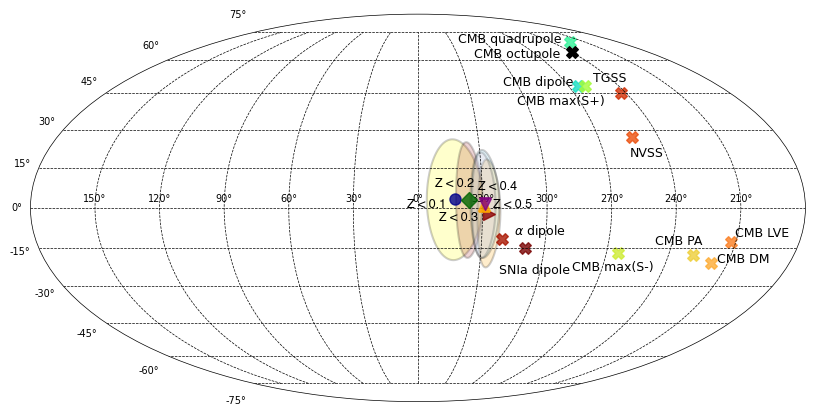}
    \caption{Mollweide projection of the cosmic preferred axis underlying the Gödel-type
    rotating model, with $1\sigma$ confidence contours, found using Pantheon+ SNIa data
    for the five redshift cutoffs considered in this work. Also shown are similar axis of
    anisotropy found
             in various astronomical/cosmological data :
             CMB dipole~\cite{plk2018maps}, CMB quadrupole and octopole axes~\cite{plk2013isostat}, CMB even and odd mirror parity axes, (hemispherical) power asymmetry (PA) axis, dipole modulation (DM) axis~\cite{plk2015isostat}, CMB PA axis from Local Variance estimator (LVE)~\cite{akrami2014}, Radio dipole in NVSS and TGSS data~\cite{bengaly2018}, SNIa Dipole~\cite{perivolaropoulos2010sn1a} and Fine structure constant ($\alpha$) dipole~\cite{alpha_dipole}.}
    \label{fig:cosmic_axes}
\end{figure}

A key parameter in this study corresponding to global rotation, $\Omega_0$,  exhibits a decreasing trend with increasing redshift bin. We obtain $\Omega_0 = 0.24 \pm 0.46$ for $Z \leq 0.1$ that essentially implies no rotation, that then peaks to $\Omega_0 = 0.29^{+0.21}_{-0.15}$ for $Z \leq 0.2$, and thereafter reducing to $\Omega_0 = 0.182 \pm 0.095$, $0.168 \pm 0.072$, and $0.120^{+0.078}_{-0.046}$ for $Z \leq 0.3$, $Z \leq 0.4$, and $Z \leq 0.5$, respectively. The trend suggests that the effects of rotation, while uncertain at low redshifts, become more stringent as higher redshift data are included.
Another dimensionless parameter `$\tilde{\rho}$', defined as $\tilde{\rho}=\sqrt{\sigma/(k+\sigma)}$, that is related to global rotation (see Eq.~\ref{eq:godel-rot-par}) shows a similar decreasing trend. Constraints derived on both $\Omega_0$ and $\tilde{\rho}$ are presented in Table~\ref{tab:cosmo-par-compare} for different redshift ranges considered. The higher the redshift bin, the smaller the value of $\tilde{\rho}$ assumes. Its values at posterior maxima varies from $\tilde{\rho}=0.0181$ to $0.0098$ from low redshift bin $Z\leq0.1$ to the highest bin $Z\leq 0.5$.

This may be indicating two possibilities. On one hand it may be suggesting that we have to go beyond the cosmographic analysis. On the other hand, rotation due to local anisotropic nature of structure could be weakening as we go farther into the Hubble flow. A fuller study of this Gödel-type model, solving the formal luminosity vs redshift relation for any redshift, could provide a resolution to this ambiguity.

The constraints on absolute SNIa magnitude `$M_0$' remain consistent being $\approx-19.2$ from the Gödel-type model as well as the $\Lambda$CDM model across all redshift ranges. Its value ranges from $M_0 \approx -19.203$ to $-19.194$. This also validates the standard candle nature of SNe Ia. Here also it is important to note that we should not compare these values directly with the Panthoen+SH0ES determination of $M_0=-19.253\pm0.027$~\cite{Riess2022SH0ES} that is estimated from the full data set. 
The astrophysical nuisance parameters $\alpha$, $\beta$, and $\gamma$ remain almost same across all redshift bins, reinforcing the robustness of SNIa light-curve corrections.
Posterior distribution of these nuisance parameters along with $\tilde{\rho}$ and $M_0$ from our MCMC analysis are shown in Fig.~[\ref{fig:1D_rho_M0_omega_by_h}].

Finally, we plot a derived parameter $\Omega_0 / h_0$ ($=\omega_0/H_0$), as a 1D contour in Fig.~\ref{fig:1D_rho_M0_omega_by_h} (bottom right panel), that captures the strength of rotation over expansion. The influence of rotation, though modest at $Z\leq 0.1$, becomes more pronounced for $Z \leq 0.2$ case where $\Omega_0 / h_0 \approx 0.29$.
Thereafter rotation to expansion ratio decrease as $\Omega_0/h_0\approx 0.182, 0.168$ and $0.120$ for $Z < 0.3, 0.4$, and $0.5$ respectively, indicating a diminishing impact of rotation as more distant SNe are included. This suggests that cosmic rotation, if present, is more relevant at lower redshifts but becomes negligible on larger scales.
Here we note that, since $h_0$ is found to be nearly same ($\sim 0.73$) across all redshift ranges, the values of $\Omega_0/h_0$ show the same trend as $\Omega_0$ parameter.

In the end, we present one of the important results of this work viz., constraints on the global axis of anisotropy in our assumed rotating cosmological model.
The inferred rotation axes are : $(R_a,D_a)\approx(251^\circ, -41^\circ)$ for $Z \leq 0.1$, $(244.8^\circ, -46^\circ)$ for $Z \leq 0.2$, $(239.6^\circ, -56^\circ)$ for $Z \leq 0.3$, $(239.3^\circ, -52^\circ)$ for $Z \leq 0.4$, and $(238.1^\circ, -52^\circ)$ for $Z \leq 0.5$ in equatorial coordinates that are also listed in Table~\ref{tab:cosmo-par-compare}.
To put these anisotropy axes constraints in a broader perspective, these axes from different redshift bins with $1\sigma$ confidence contours are shown in galactic coordinates along with other anisotropy axes seen in various astronomical and cosmological data sets.

One of the earliest observational constraints on cosmic rotation was reported by Birch in 1982 where a dipole anisotropy in the distribution of observed angle between polarization and position angles of 137
radio sources was interpreted as evidence for cosmic rotation~\cite{birch1982}. Birch reported the cosmic rotation axis as $(R_a,D_a)\approx(200^\circ, -40^\circ)$ in equatorial coordinates. Later, the same was confirmed using `indirectional statistics' by Kendall and Young~\cite{KendallYoung1984} with a higher significance who reported the axis of rotation to be $(R_a,D_a)\approx(202^\circ,-37^\circ)$ (using a sample of 134 sources) which is essentially same as that reported by Birch.
Using a different sample of 205 sources, Andreasyan reported a cosmic anisotropy in the direction $(R_a,D_a)\approx(188^\circ,-11^\circ)$~\cite{Andreasyan1986}.
Considering a subsample of Birch's data (with 51 sources whose redshifts were known), a global axis of rotation within the present framework of the Gödel-like model was found to be $(R_a,D_a)\approx(184^\circ,-38^\circ)$~\cite{1992obukhov}.

The present model of our interest given by Eq.~(\ref{eq:go_metric}) was earlier analyzed in Ref.~\cite{Jain2007Godel} using the \emph{gold} and \emph{silver} Type Ia supernovae data sets of Ref.~\cite{Riess2004}. However, in that study, the rotation term (with the factor $\Omega_0/h_0$) in Eq.~(\ref{eq:godel_distance_modulus}) was ignored constraining only the anisotropy axis (of rotation) and the parameter $\tilde{\rho}$. The \emph{gold} data set had 157 SNIa and \emph{silver} data set contained 29 sources. Because of the small sample size, perhaps, they could only find an upper bound on $\tilde{\rho}$ as $<0.049$ and $<0.046$, and the axis of anisotropy was found to be $(R_a,D_a)\approx(229.5^\circ,-53.2^\circ)$ and $(199.6^\circ,-47.5^\circ)$ in equatorial coordinates, when using the \emph{gold} and \emph{gold}+\emph{silver} data sets respectively.

Future observations, particularly with additional intermediate and low redshift supernovae, will help refine these constraints and further assess the robustness of the anisotropy signal.

\subsection*{Model comparison through AIC}
\label{sec:aic-criteria}

To qualitatively parameterize a model, one must account for the number of parameters used in model fitting. Since introducing more free parameters in a model usually results in a better fit, we should therefore have a mechanism to either justify or penalize these additional free parameters introduced in a model. To compare different models with different numbers of parameters, it is essential to balance the fit quality against the added complexity. A widely used tool for this is the Akaike Information Criterion (AIC), denoted here as `$\mathcal{A}$'. Given two models, the preferred model is the one with the minimum AIC value. Note that the AIC value for a model itself cannot define the quality/goodness of fit. It is only the relative AIC values that are used to judge the goodness of fit of one model with respect to another model in describing the data at hand. Based on principles of information theory, `$\mathcal{A}$' penalizes model complexity to avoid overfitting while favoring those that effectively/efficiently describe the data~\cite{AIC1974}.

It is defined as,
\begin{equation}
\mathcal{A} = 2k - 2\ln(\mathcal{L_{\max}}) = 2k + \chi^2_{\rm min}\,,
\end{equation}
where `$k$' is the number of parameters, $\mathcal{L_{\max}}$ is the maximum likelihood for the given model and $\chi^2_{\rm min}$ is the minimum of the $\chi^2$ function (Eq.~\ref{eq:chi2-cov}) in the parameter space after optimization. The term `$2k$' penalizes models with higher complexity. To quantify the relative performance of different models, difference in the values of $\mathcal{A}$ are computed with respect to a reference model as,
\begin{equation}
\Delta\mathcal{A}_i = \mathcal{A}_{\text{model},i} - \mathcal{A}_{\text{ref}}\,,
\end{equation}
where $\Delta\mathcal{A}_i$ is the relative AIC value of $i$-th model compared to the \emph{reference} model. A lower value of $\Delta\mathcal{A}_i$ indicates that the $i$-th model describes the data better from among all candidate models. Additionally, the probability of each model relative to the reference model is given by~\cite{Burnham_04},
\begin{equation}
\mathcal{P}_i = \exp\left(-\frac{\Delta\mathcal{A}_i}{2}\right)\,.
\end{equation}

In this study, the flat $\Lambda$CDM model serves as the reference, and the values of $\Delta\mathcal{A}$ and the corresponding probabilities $\mathcal{P}$ are used to evaluate the strength of support for each candidate model. Smaller $\Delta\mathcal{A}$ values indicate stronger evidence for a model, with $\Delta\mathcal{A} \leq 2$ representing strong support, values in the range $4 \leq \Delta\mathcal{A} \leq 7$ suggesting moderate to weak support, and $\Delta\mathcal{A} \geq 10$ indicating no support~\cite{Burnham_04}. This criterion allows for a quantitative comparison of the likelihood of different models in describing the observed data.

The AIC ($\mathcal{A}$), relative AIC ($\Delta\mathcal{A}$) and the relative model probabilities  ($\mathcal{P}_i$) for our rotating Gödel-type model as well as the standard $\Lambda$CDM model for all redshift bins considered are tabulated in Table~\ref{tab:cosmo-par-compare}.
The interpretation provided by these AIC values~\cite{AIC1974,Burnham_04} shows that the Gödel's rotating model is either competitive or strongly preferred over the base (reference) flat $\Lambda$CDM model for all choices of redshift range considered in this cosmographic analysis. Specifically, for $Z \leq 0.1$, the rotating model is almost equally ($\sim$ 1.1 times) probable, while it is $\sim$ 3 , 10, 4.8 and  6.9 times more probable compared to the base model for the redshift cutoffs $Z\leq 0.2, 0.3, 0.4$, and 0.5, respectively. 

This result is unexpected and is interesting, since the preference of rotating model over the standard model, particularly at $Z\geq0.1$, suggests that the additional kinematic effects introduced by rotation might play a non-negligible role in the late-time expansion history of the universe. Although the statistical evidence favors the rotating model in this analysis, it is important to interpret these results cautiously since the significance of this preference depends on the robustness of the cosmographic expansion employed and the impact of potential systematics in the data. Additional observational constraints, such as those including energy density fractions by deriving the Friedman equations in Gödel type rotating solution, may provide further insights into whether this preference is a genuine physical effect or a statistical anomaly.

\section{Conclusions}
\label{sec:conclusion}

Rotation is a generic feature arising in any astrophysical setting that is an observational fact as seen in a wide variety of celestial systems.
In this study, we explored the possibility of cosmic rotation for our Universe using a Gödel-type metric and observational data from Type Ia supernovae across multiple redshift ranges. We performed a cosmographic analysis of the magnitude vs redshift ($m-Z$) relation derived using the Kristian-Sachs formalism. We constrained both cosmological and astrophysical parameters using an MCMC method. Our results indicate interesting deviations from standard $\Lambda$CDM constraints for various redshift bins, particularly in the derived parameter $\Omega_0/h_0(=\omega_0/H_0)$, that captures the strength of rotation vs expansion.

The inferred cosmological parameters, particularly the (dimensionless) Hubble constant $h_0$ and the deceleration parameter $q_0$, are central to understanding the effects of rotation in our analysis. We find that $h_0 \sim 0.73$ being virtually same for all choices of redshift bins, indicating that the inclusion of rotational degrees of freedom in the Gödel-type model does not significantly alter the overall expansion rate compared to \(\Lambda\)CDM. However, the deceleration parameter exhibits notable variations across redshift bins, with values ranging from \( q_0 \approx 0.01 \) at \( Z \leq 0.1 \) to \( q_0 \approx -0.089 \) at \( Z \leq 0.4 \). In contrast, $q_0 = -0.51$ for the standard flat \(\Lambda\)CDM model from Pantheon+ SNe Ia constraints~\cite{Riess2022SH0ES}. The small positive value at very low redshift and the increasingly negative values at higher redshifts may arise due to statistical limitations in the dataset or subtle rotational effects influencing local expansion. As more high-redshift supernovae are included, \( q_0 \) approaches the standard \(\Lambda\)CDM expectation, albeit with slightly less negative values than those obtained in a purely isotropic model. This suggests that while rotation does not drastically impact cosmic acceleration, it may introduce small but noticeable deviations from the standard paradigm. 

To quantify the relative impact of rotation versus expansion, we analyzed the dimensionless parameter \( \Omega_0 / h_0 \), which provides insight into the strength of cosmic rotation across different redshift bins. Our results indicate that rotational effects are most pronounced at \( Z \geq 0.1 \), reaching a maximum value of $\approx0.40$ for the \( Z \leq 0.2 \) bin. At lower redshifts (\( Z < 0.1 \)), this ratio is relatively small (essentially consistent with \emph{zero} within error bars). However, as we include higher redshifts, the influence of rotation is found to diminish steadily, with $\Omega_0/h_0$ decreasing to \( \approx 0.16 \) for \( Z \leq 0.5 \) bin. This decline suggests that rotational effects, if present, are confined to relatively local cosmic scales and become negligible when considering more distant supernovae.  These results highlight that while rotation may play a role in the cosmic evolution at small redshifts, its contribution becomes increasingly negligible on larger scales, reinforcing the overall isotropy of the universe. The global rotation axis inferred from our model is consistent across all redshift bins, with its mean direction given by \((R_a,D_a) \approx (243^\circ, -49^\circ) \).

A related parameter that captures global rotation in this model is $\tilde{\rho}=\sqrt{\sigma/(k+\sigma)}$. It follows the same trend as $q_0$ and $\Omega_0$ (and hence $\Omega_0/h_0$). It values decreases steadily from the lowest ($Z\leq0.1$) to highest ($Z\leq0.5$) redshift ranges chosen in the present study. Astrophysical parameters viz., $M_0$, $\alpha$, $\beta$, and $\gamma$, which characterize supernova magnitude calibration, its light-curve stretch, color and host galaxy properties, are nearly same across different redshift ranges. This reinforces the robustness of the dataset and suggests that systematic effects from supernova standardization are minimal.

The AIC-based model comparison reveals an intriguing preference for the Gödel-type rotating model over the standard flat $\Lambda$CDM cosmology, particularly at higher redshifts. While both models remain competitive for $Z \leq 0.1$, with the candidate model being only marginally more probable, the preference for the rotating model increases significantly at larger redshift cutoffs. Specifically, for $Z \leq 0.3$, $0.4$, and $0.5$, the Gödel-type model is found to be $\sim 9.9$, $4.7$, and $6.9$ times more probable than $\Lambda$CDM model, respectively. This suggests that the additional kinematic effects introduced by cosmic rotation might play a non-negligible role in the late-time expansion history. However, caution is warranted in interpreting these results. The robustness of the cosmographic approach and potential systematics unaccounted for in the data could have a bearing on these findings and further investigation - fully testing this model going beyond the cosmographic expansion - are necessary to determine whether this preference is indicative of a true physical effect or a statistical fluke or a data artifact.

In conclusion, our analysis shows indications for cosmic rotation at intermediate reshifts given that the AIC model assessment metric strongly favours the Gödel-type model over the standard model. Nevertheless it needs further study to derive better constraints on the model to fully understand the ramifications of cosmic rotation more clearly. Future observations from upcoming surveys like Euclid~\cite{euclid_collaboration} and JWST~\cite{jwst_collaboration} may provide improved constraints on both the rotation parameter and the anisotropy axis. This could offer new insights into cosmic isotropy and potential deviations that enrich our understanding of the universe's large-scale structure.

\section*{Acknowledgements}
AV acknowledges the financial support received through a research fellowship awarded by Council of Scientific \& Industrial Research (CSIR), India during this project. We acknowledge National Supercomputing Mission (NSM) for providing computing resources of ‘PARAM Shivay’ at Indian Institute of Technology (BHU), Varanasi, which is implemented by C-DAC and supported by the Ministry of Electronics and Information Technology (MeitY) and Department of Science and Technology (DST), Government of India.
 DFM thanks the Research Council of Norway for their support and the resources provided by UNINETT Sigma2 -- the National Infrastructure for High-Performance Computing and Data Storage in Norway.
We acknowledge using \texttt{Cobaya}\footnote{\url{https://cobaya.readthedocs.io/en/latest/}}  \cite{torrado2021} - a code for Bayesian analysis in Cosmology, and \texttt{GetDist}\footnote{\url{https://getdist.readthedocs.io/en/latest/}} \cite{lewis:2019xzd} - a user friendly GUI for the analysis and plotting of MCMC samples. We also acknowledge the use of \texttt{SciPy}\footnote{\url{https://scipy.org/}}~\cite{scipy2020}, \texttt{NumPy}\footnote{\url{https://numpy.org/}}~\cite{numpy2020}, \texttt{Astropy}\footnote{\url{http://www.astropy.org}}~\cite{astropy2013,astropy2018,astropy2022}, and \texttt{Matplotlib}\footnote{\url{https://matplotlib.org/}}~\cite{hunter:2007}.

\bibliographystyle{unsrt}
\bibliography{ref_godel}

\begin{thebibliography}{10}

\bibitem{1945whittaker}
E.T. {Whittaker}.
\newblock {Spin in the Universe}.
\newblock {\em Yearbook of Royal Society}, Edinburgh:513, 1945.

\bibitem{1948lemaitre_atom_hypothesis}
G.~{Lemâıtre}.
\newblock {L’hypothèse de l’atome primitif}.
\newblock {\em Acta Pontifica Acad. Sci.}, 12:25--40, 1948.

\bibitem{1924lanczos}
K.~{Lanczos}.
\newblock {Über eine stationäre Kosmologie im Sinne der Einsteinschen
  Gravitationstheorie}.
\newblock {\em Zeitschrift für Physik}, 21:73--110, 1924.
\newblock [English translation: On a Stationary cosmology in the sense of
  Einstein’s theory of gravitation, Gen. Rel. Grav. 29 (1997) 361-399].

\bibitem{1946gamow}
G.~{Gamow}.
\newblock {Rotating universe?}
\newblock {\em Nature}, 158:549, 1946.

\bibitem{1949godel}
K.~{Gödel}.
\newblock {An example of a new type of cosmological solutions of Einstein’s
  field equations of gravitation}.
\newblock {\em Reviews of Modern Physics}, 21:447--450, 1949.

\bibitem{1936stockum}
W.J. {van Stockum}.
\newblock {Gravitational field of a rotating mass as an example of
  algebraically special metrics}.
\newblock {\em Proceedings of the Royal Society of Edinburgh}, 57:135--154,
  1936/37.

\bibitem{1952godel}
K.~{Gödel}.
\newblock {Rotating universes in general relativity theory}.
\newblock In {\em Proceedings of the International Congress of Mathematicians},
  volume~1, pages 175--181, Cambridge, USA, 1952.

\bibitem{1966maitra}
S.~{Maitra}.
\newblock {Some solutions of Einstein’s field equations}.
\newblock {\em Journal of Mathematical Physics}, 7:1025, 1966.

\bibitem{1969ozsvath}
I.~{Ozsvath} and E.~{Schücking}.
\newblock {An anti-Mach metric}.
\newblock {\em Annals of Physics}, 55:166, 1969.

\bibitem{1985bedran_cosmology_torsion}
M.~{Bedran}, V.~{Vaidya}, and M.~{Som}.
\newblock {Stationary cosmological solutions with torsion}.
\newblock {\em Nuovo Cimento B}, 87:101--108, 1985.

\bibitem{1985smalley_godel_cosmology}
L.L. {Smalley}.
\newblock {Gödel cosmology in Riemann-Cartan space-time with spin density}.
\newblock {\em Physical Review D}, 32:3124--3127, 1985.

\bibitem{1986smalley_self_consistent_godel_cosmology}
L.L. {Smalley}.
\newblock {Self-consistent Gödel cosmology with spin density in Riemann-Cartan
  space-time}.
\newblock {\em Physics Letters A}, 113:463--466, 1986.

\bibitem{1986duarte_homogeneous_cosmos}
J.~{Duarte de Oliveira}, A.F.F. {Teixeira}, and J.~{Tiomno}.
\newblock {Homogeneous cosmos of Weyssenhoff fluid in Einstein-Cartan space}.
\newblock {\em Physical Review D}, 34:3661--3665, 1986.

\bibitem{1987obukhov_weyssenhoff_fluid}
Yu.N. {Obukhov} and V.A. {Korotky}.
\newblock {The Weyssenhoff fluid in Einstein-Cartan theory}.
\newblock {\em Classical and Quantum Gravity}, 4:1633--1657, 1987.

\bibitem{1988pavelkin_rotation_cosmology}
V.N. {Pavelkin} and V.F. {Panov}.
\newblock {A study of rotation in cosmology}.
\newblock {\em Sov. Phys. J.}, 31:541--544, 1989.

\bibitem{1992korotky_obukhov_ecsk}
V.A. {Korotky} and Yu.N. {Obukhov}.
\newblock {Rotating and expanding cosmology in ECSK-theory}.
\newblock {\em Astrophysics and Space Science}, 198:1--12, 1992.

\bibitem{1995palle_imperfect_fluid}
D.~{Palle}.
\newblock {Imperfect fluid in the Einstein-Cartan gravity}.
\newblock {\em Preprint Dept. Theor. Phys., R. Boskovic Inst.}, 1995.
\newblock 6 p.

\bibitem{1996palle_cosmological_observables}
D.~{Palle}.
\newblock {On certain relationships between cosmological observables in the
  Einstein-Cartan gravity}.
\newblock {\em Nuovo Cimento B}, 111:671--675, 1996.

\bibitem{1998aman_riemann_cartan_space_times}
J.E. {Åman}, J.B. {Fonseca-Neto}, M.A.H. {MacCallum}, and M.J. {Rebouças}.
\newblock {Riemann-Cartan space-times of Gödel type}.
\newblock {\em Classical and Quantum Gravity}, 15:1089--1101, 1998.

\bibitem{maccallum1969}
G.~F.~R. {Ellis} and M.~A.~H. {MacCallum}.
\newblock {A class of homogeneous cosmological models}.
\newblock {\em Communications in Mathematical Physics}, 12(2):108--141, June
  1969.

\bibitem{maccallum1970}
M.~A.~H. {MacCallum} and G.~F.~R. {Ellis}.
\newblock {A class of homogeneous cosmological models: II. Observations}.
\newblock {\em Communications in Mathematical Physics}, 19(1):31--64, March
  1970.

\bibitem{maccallum1971}
M.~A.~H. {MacCallum}.
\newblock {A class of homogeneous cosmological models III: Asymptotic
  behaviour}.
\newblock {\em Communications in Mathematical Physics}, 20(1):57--84, March
  1971.

\bibitem{1983barrow_nature}
J.D. {Barrow}, R.~{Juszkiewicz}, and D.H. {Sonoda}.
\newblock {Structure of the cosmic microwave background}.
\newblock {\em Nature}, 305:397--402, 1983.

\bibitem{1985barrow_mnras}
J.D. {Barrow}, R.~{Juszkiewicz}, and D.H. {Sonoda}.
\newblock {Universal rotation: how large can it be?}
\newblock {\em Monthly Notices of the Royal Astronomical Society},
  213:917--943, 1985.

\bibitem{1969hawking_mnras}
S.W. {Hawking}.
\newblock {On the rotation of the universe}.
\newblock {\em Monthly Notices of the Royal Astronomical Society},
  142:129--141, 1969.

\bibitem{1973collins_hawking}
C.B. {Collins} and S.W. {Hawking}.
\newblock {The rotation and distortion of the universe}.
\newblock {\em Monthly Notices of the Royal Astronomical Society},
  162:307--320, 1973.

\bibitem{1997kogut_prd}
A.~{Kogut}, G.~{Hinshaw}, and A.J. {Banday}.
\newblock {Limits to global rotation and shear from the COBE DMR four year sky
  maps}.
\newblock {\em Physical Review D}, 55:1901--1905, 1997.

\bibitem{1970wolfe_apj}
A.M. {Wolfe}.
\newblock {New limits on the shear and rotation of the universe from the X-ray
  background}.
\newblock {\em Astrophysical Journal}, 159:L61--L67, 1970.

\bibitem{1990obukhov}
Yuri~N. {Obukhov}.
\newblock {Gauge Theories of Fundamental Interactions: Proceedings of the XXXII
  Semester of the S. Banach International Mathematical Center}.
\newblock In M.~{Pawlowski} and R.~{Raczka}, editors, {\em Proceedings of the
  XXXII Semester of the S. Banach Internat. Math. Center, Warsaw, Poland, 19
  Sept-3 Dec 1988}, page 341, Singapore, 1990. World Scientific.

\bibitem{1992obukhov}
Yu.~N. {Obukhov}.
\newblock {Rotation in cosmology}.
\newblock {\em General Relativity and Gravitation}, 24(2):121--128, February
  1992.

\bibitem{birch1982}
P.~{Birch}.
\newblock {Is the Universe rotating?}
\newblock {\em \nat}, 298(5873):451--454, July 1982.

\bibitem{1982birch_reply}
P.~{Birch}.
\newblock {Is there evidence for universal rotation? Birch replies}.
\newblock {\em Nature}, 301:736, 1982.

\bibitem{1983conway_mnras}
R.G. {Conway}, P.~{Birch}, R.J. {Davis}, L.R. {Jones}, A.J. {Kerr}, and
  D.~{Stannard}.
\newblock {Multi-frequency observations of linear polarization in 94 sources
  from the 3CR catalogue. I. Observations}.
\newblock {\em Monthly Notices of the Royal Astronomical Society},
  202:813--823, 1983.

\bibitem{aluri2023cp}
Pavan {Kumar Aluri}, Paolo {Cea}, Pravabati {Chingangbam}, Ming-Chung {Chu},
  Roger~G. {Clowes}, Damien {Hutsem{\'e}kers}, Joby~P. {Kochappan}, Alexia~M.
  {Lopez}, Lang {Liu}, Niels C.~M. {Martens}, C.~J.~A.~P. {Martins},
  Konstantinos {Migkas}, Eoin {{\'O} Colg{\'a}in}, Pratyush {Pranav}, Lior
  {Shamir}, Ashok~K. {Singal}, M.~M. {Sheikh-Jabbari}, Jenny {Wagner},
  Shao-Jiang {Wang}, David~L. {Wiltshire}, Shek {Yeung}, Lu~{Yin}, and Wen
  {Zhao}.
\newblock {Is the observable Universe consistent with the cosmological
  principle?}
\newblock {\em Classical and Quantum Gravity}, 40(9):094001, May 2023.

\bibitem{wmap7yranom}
C.~L. {Bennett}, R.~S. {Hill}, G.~{Hinshaw}, D.~{Larson}, K.~M. {Smith},
  J.~{Dunkley}, B.~{Gold}, M.~{Halpern}, N.~{Jarosik}, A.~{Kogut},
  E.~{Komatsu}, M.~{Limon}, S.~S. {Meyer}, M.~R. {Nolta}, N.~{Odegard},
  L.~{Page}, D.~N. {Spergel}, G.~S. {Tucker}, J.~L. {Weiland}, E.~{Wollack},
  and E.~L. {Wright}.
\newblock {Seven-year Wilkinson Microwave Anisotropy Probe (WMAP) Observations:
  Are There Cosmic Microwave Background Anomalies?}
\newblock {\em \apjs}, 192(2):17, February 2011.

\bibitem{wmap9yrmaps}
C.~L. {Bennett}, D.~{Larson}, J.~L. {Weiland}, N.~{Jarosik}, G.~{Hinshaw},
  N.~{Odegard}, K.~M. {Smith}, R.~S. {Hill}, B.~{Gold}, M.~{Halpern},
  E.~{Komatsu}, M.~R. {Nolta}, L.~{Page}, D.~N. {Spergel}, E.~{Wollack},
  J.~{Dunkley}, A.~{Kogut}, M.~{Limon}, S.~S. {Meyer}, G.~S. {Tucker}, and
  E.~L. {Wright}.
\newblock {Nine-year Wilkinson Microwave Anisotropy Probe (WMAP) Observations:
  Final Maps and Results}.
\newblock {\em \apjs}, 208(2):20, October 2013.

\bibitem{plk2013isostat}
P.~A.~R. Ade et~al.
\newblock {Planck 2013 results. XXIII. Isotropy and statistics of the CMB}.
\newblock {\em Astron. Astrophys.}, 571:A23, 2014.

\bibitem{plk2015isostat}
P.~A.~R. Ade et~al.
\newblock {Planck 2015 results. XVI. Isotropy and statistics of the CMB}.
\newblock {\em Astron. Astrophys.}, 594:A16, 2016.

\bibitem{plk2018isostat}
Y.~Akrami et~al.
\newblock {Planck 2018 results. VII. Isotropy and Statistics of the CMB}.
\newblock {\em Astron. Astrophys.}, 641:A7, 2020.

\bibitem{johnjain2004}
John~P. {Ralston} and Pankaj {Jain}.
\newblock {The Virgo Alignment Puzzle in Propagation of Radiation on
  Cosmological Scales}.
\newblock {\em International Journal of Modern Physics D}, 13(9):1857--1877,
  January 2004.

\bibitem{bull2016}
Philip Bull et~al.
\newblock {Beyond {\ensuremath{\Lambda}} CDM: Problems, solutions, and the road
  ahead}.
\newblock {\em Phys. Dark Univ.}, 12:56--99, 2016.

\bibitem{abdalla2022}
Elcio Abdalla et~al.
\newblock {Cosmology intertwined: A review of the particle physics,
  astrophysics, and cosmology associated with the cosmological tensions and
  anomalies}.
\newblock {\em JHEAp}, 34:49--211, 2022.

\bibitem{perivolaropoulos2022}
L.~{Perivolaropoulos} and F.~{Skara}.
\newblock {Challenges for {\ensuremath{\Lambda}}CDM: An update}.
\newblock {\em New Astron. Rev.}, 95:101659, December 2022.

\bibitem{Efstathiou:2020}
G.~{Efstathiou}.
\newblock {A Lockdown Perspective on the Hubble Tension (with comments from the
  SH0ES team)}.
\newblock {\em arXiv e-prints}, page arXiv:2007.10716, July 2020.

\bibitem{valentino:2021}
Eleonora {Di Valentino}, Olga {Mena}, Supriya {Pan}, Luca {Visinelli}, Weiqiang
  {Yang}, Alessandro {Melchiorri}, David~F. {Mota}, Adam~G. {Riess}, and Joseph
  {Silk}.
\newblock {In the realm of the Hubble tension-a review of solutions}.
\newblock {\em Classical and Quantum Gravity}, 38(15):153001, July 2021.

\bibitem{freedman2021}
Wendy~L. {Freedman}.
\newblock {Measurements of the Hubble Constant: Tensions in Perspective}.
\newblock {\em \apj}, 919(1):16, September 2021.

\bibitem{2023Kamionkowski}
Marc {Kamionkowski} and Adam~G. {Riess}.
\newblock {The Hubble Tension and Early Dark Energy}.
\newblock {\em Annual Review of Nuclear and Particle Science}, 73:153--180,
  September 2023.

\bibitem{2021Xia_MNRAS}
Qianli {Xia}, Mark~C. {Neyrinck}, Yan-Chuan {Cai}, and Miguel~A.
  {Arag{\'o}n-Calvo}.
\newblock {Intergalactic filaments spin}.
\newblock {\em \mnras}, 506(1):1059--1072, September 2021.

\bibitem{2022Alexander}
Stephon {Alexander}, Christian {Capanelli}, Elisa {G.~M. Ferreira}, and Evan
  {McDonough}.
\newblock {Cosmic filament spin from dark matter vortices}.
\newblock {\em Physics Letters B}, 833:137298, October 2022.

\bibitem{Shamir2012}
Lior {Shamir}.
\newblock {Handedness asymmetry of spiral galaxies with $z<0.3$ shows cosmic
  parity violation and a dipole axis}.
\newblock {\em Physics Letters B}, 715(1):25--29, August 2012.

\bibitem{Shamir2020}
Lior {Shamir}.
\newblock {Large‑scale asymmetry between clockwise and counterclockwise
  galaxies revisited}.
\newblock {\em Astronomische Nachrichten}, 341(3):324--330, March 2020.

\bibitem{Shamir2023}
Lior {Shamir}.
\newblock {Large-Scale Asymmetry in the Distribution of Galaxy Spin
  Directions{\textemdash}Analysis and Reproduction}.
\newblock {\em Symmetry}, 15(9):1704, September 2023.

\bibitem{2019Lee}
Joon~Hyeop {Lee}, Mina {Pak}, Hye-Ran {Lee}, and Hyunmi {Song}.
\newblock {Galaxy Rotation Coherent with the Motions of Neighbors: Discovery of
  Observational Evidence}.
\newblock {\em \apj}, 872(1):78, February 2019.

\bibitem{Lee2019}
Joon~Hyeop {Lee}, Mina {Pak}, Hyunmi {Song}, Hye-Ran {Lee}, Suk {Kim}, and
  Hyunjin {Jeong}.
\newblock {Mysterious Coherence in Several-megaparsec Scales between Galaxy
  Rotation and Neighbor Motion}.
\newblock {\em \apj}, 884(2):104, October 2019.

\bibitem{1972ApJ_Silk}
Joseph {Silk} and Susan {Ames}.
\newblock {Primordial Turbulence and the FORMAT10N of Galaxies}.
\newblock {\em \apj}, 178:77--94, November 1972.

\bibitem{1996Chernin}
Arthur~D. {Chernin}.
\newblock {Shocks and vorticity in cosmic hydrodynamics}.
\newblock {\em Vistas in Astronomy}, 40(2):257--301, January 1996.

\bibitem{2009Pontzen}
Andrew {Pontzen}.
\newblock {Rogues' gallery: The full freedom of the Bianchi CMB anomalies}.
\newblock {\em \prd}, 79(10):103518, May 2009.

\bibitem{coles2011}
Rockhee {Sung} and Peter {Coles}.
\newblock {Temperature and polarization patterns in anisotropic cosmologies}.
\newblock {\em \jcap}, 2011(6):036, June 2011.

\bibitem{Jaffe2005ApJ}
T.~R. {Jaffe}, A.~J. {Banday}, H.~K. {Eriksen}, K.~M. {G{\'o}rski}, and F.~K.
  {Hansen}.
\newblock {Evidence of Vorticity and Shear at Large Angular Scales in the WMAP
  Data: A Violation of Cosmological Isotropy?}
\newblock {\em \apjl}, 629(1):L1--L4, August 2005.

\bibitem{jaffe2006}
T.~R. {Jaffe}, S.~{Hervik}, A.~J. {Banday}, and K.~M. {G{\'o}rski}.
\newblock {On the Viability of Bianchi Type VII$_{h}$ Models with Dark Energy}.
\newblock {\em \apj}, 644(2):701--708, June 2006.

\bibitem{saadeh2016MNRAS}
Daniela {Saadeh}, Stephen~M. {Feeney}, Andrew {Pontzen}, Hiranya~V. {Peiris},
  and Jason~D. {McEwen}.
\newblock {A framework for testing isotropy with the cosmic microwave
  background}.
\newblock {\em \mnras}, 462(2):1802--1811, October 2016.

\bibitem{1991KorotkyObukhov}
V.~A. {Korotkij} and Yu.~N. {Obukhov}.
\newblock {Kinematic analysis of cosmological models with rotation.}
\newblock {\em Soviet Journal of Experimental and Theoretical Physics},
  72(1):11--15, January 1991.
\newblock [ZhETF 99 (1991) 22, in Russian].

\bibitem{1996obukhov}
Vladimir~A. {Korotky} and Yuri~N. {Obukhov}.
\newblock {On Cosmic Rotation}.
\newblock In P.~{Pronin} and G.~{Sardanashvily}, editors, {\em Gravity
  Particles and Space-time}, pages 421--439. World Scientific, Singapore,
  January 1996.

\bibitem{2000obukhov}
Yuri~N. {Obukhov}.
\newblock {On physical foundations and observational effects of cosmic
  rotation}.
\newblock In M.~{Scherfner}, T.~{Chrobok}, and M.~{Shefaat}, editors, {\em
  Colloquium on Cosmic Rotation}, page~23. Wissenschaft und Technik Verlag,
  Berlin, January 2000.

\bibitem{1966Kristian_Sachs_ApJ}
J.~{Kristian} and R.~K. {Sachs}.
\newblock {Observations in Cosmology}.
\newblock {\em \apj}, 143:379, February 1966.

\bibitem{scolnic2022}
Dan Scolnic et~al.
\newblock {The Pantheon+ Analysis: The Full Data Set and Light-curve Release}.
\newblock {\em Astrophys. J.}, 938(2):113, 2022.

\bibitem{Leavitt:1908vb}
Henrietta~S. Leavitt.
\newblock {1777 variables in the Magellanic Clouds}.
\newblock {\em Harvard Obs. Annals}, 60:87--108, 1908.

\bibitem{Leavitt1912}
Henrietta~S. {Leavitt} and Edward~C. {Pickering}.
\newblock {Periods of 25 Variable Stars in the Small Magellanic Cloud.}
\newblock {\em Harvard College Observatory Circular}, 173:1--3, March 1912.

\bibitem{Riess2022SH0ES}
Adam~G. Riess et~al.
\newblock {A Comprehensive Measurement of the Local Value of the Hubble
  Constant with {1 km s$^{-1}$ Mpc$^{-1}$} Uncertainty from the Hubble Space
  Telescope and the SH0ES Team}.
\newblock {\em Astrophys. J. Lett.}, 934(1):L7, 2022.

\bibitem{Brout2022Pantheonplus}
Dillon Brout et~al.
\newblock {The Pantheon+ Analysis: Cosmological Constraints}.
\newblock {\em Astrophys. J.}, 938(2):110, 2022.

\bibitem{plk2018maps}
N.~Aghanim et~al.
\newblock {Planck 2018 results. I. Overview and the cosmological legacy of
  Planck}.
\newblock {\em Astron. Astrophys.}, 641:A1, 2020.

\bibitem{akrami2014}
Y.~{Akrami}, Y.~{Fantaye}, A.~{Shafieloo}, H.~K. {Eriksen}, F.~K. {Hansen},
  A.~J. {Banday}, and K.~M. {G{\'o}rski}.
\newblock {Power Asymmetry in WMAP and Planck Temperature Sky Maps as Measured
  by a Local Variance Estimator}.
\newblock {\em \apjl}, 784(2):L42, April 2014.

\bibitem{bengaly2018}
Carlos A.~P. {Bengaly}, Roy {Maartens}, and Mario~G. {Santos}.
\newblock {Probing the Cosmological Principle in the counts of radio galaxies
  at different frequencies}.
\newblock {\em \jcap}, 2018(4):031, April 2018.

\bibitem{perivolaropoulos2010sn1a}
I.~{Antoniou} and L.~{Perivolaropoulos}.
\newblock {Searching for a cosmological preferred axis: Union2 data analysis
  and comparison with other probes}.
\newblock {\em \jcap}, 2010(12):012, December 2010.

\bibitem{alpha_dipole}
Antonio {Mariano} and Leandros {Perivolaropoulos}.
\newblock {Is there correlation between fine structure and dark energy cosmic
  dipoles?}
\newblock {\em \prd}, 86(8):083517, October 2012.

\bibitem{KendallYoung1984}
D.~G. {Kendall} and G.~A. {Young}.
\newblock {Indirectional statistics and the significance of an asymmetry
  discovered by Birch}.
\newblock {\em \mnras}, 207:637--647, April 1984.

\bibitem{Andreasyan1986}
R.~R. {Andreasyan}.
\newblock {The problem of the existence of large-scale anisotropy in
  metagalactic space}.
\newblock {\em Astrophysics}, 24(2):213--220, March 1986.

\bibitem{Jain2007Godel}
Pankaj {Jain}, Moninder~S. {Modgil}, and John~P. {Ralston}.
\newblock {Search for Global Metric Anisotropy in Type 1a Supernova Data}.
\newblock {\em Modern Physics Letters A}, 22(16):1153--1165, January 2007.

\bibitem{Riess2004}
Adam~G. {Riess} et~al.
\newblock {Type Ia Supernova Discoveries at $z > 1$ from the Hubble Space
  Telescope: Evidence for Past Deceleration and Constraints on Dark Energy
  Evolution}.
\newblock {\em \apj}, 607(2):665--687, June 2004.

\bibitem{AIC1974}
H.~{Akaike}.
\newblock {A New Look at the Statistical Model Identification}.
\newblock {\em IEEE Transactions on Automatic Control}, 19:716--723, January
  1974.

\bibitem{Burnham_04}
K.~P. Burnham and D.~R. Anderson.
\newblock Multimodel inference: {Understanding} {AIC} and {BIC} in model
  selection.
\newblock {\em Sociological Methods and Research}, 33(2):261--304, November
  2004.

\bibitem{euclid_collaboration}
Y.~Mellier et~al.
\newblock {Euclid. I. Overview of the Euclid mission}.
\newblock {\em Astron. Astrophys.}, 697:A1, 2025.

\bibitem{jwst_collaboration}
Jonathan~P. {Gardner} et~al.
\newblock {The James Webb Space Telescope Mission}.
\newblock {\em \pasp}, 135(1048):068001, June 2023.

\bibitem{torrado2021}
Jes{\'u}s {Torrado} and Antony {Lewis}.
\newblock {Cobaya: code for Bayesian analysis of hierarchical physical models}.
\newblock {\em \jcap}, 2021(5):057, May 2021.

\bibitem{lewis:2019xzd}
Antony {Lewis}.
\newblock {GetDist: a Python package for analysing Monte Carlo samples}.
\newblock {\em arXiv e-prints}, page arXiv:1910.13970, October 2019.

\bibitem{scipy2020}
Pauli Virtanen et~al.
\newblock {SciPy 1.0--Fundamental Algorithms for Scientific Computing in
  Python}.
\newblock {\em Nature Meth.}, 17:261, 2020.

\bibitem{numpy2020}
Charles~R. Harris et~al.
\newblock {Array programming with NumPy}.
\newblock {\em Nature}, 585(7825):357--362, 2020.

\bibitem{astropy2013}
Thomas~P. Robitaille et~al.
\newblock {Astropy: A Community Python Package for Astronomy}.
\newblock {\em Astron. Astrophys.}, 558:A33, 2013.

\bibitem{astropy2018}
A.~M. Price-Whelan et~al.
\newblock {The Astropy Project: Building an Open-science Project and Status of
  the v2.0 Core Package}.
\newblock {\em Astron. J.}, 156(3):123, 2018.

\bibitem{astropy2022}
Adrian~M. Price-Whelan et~al.
\newblock {The Astropy Project: Sustaining and Growing a Community-oriented
  Open-source Project and the Latest Major Release (v5.0) of the Core
  Package*}.
\newblock {\em Astrophys. J.}, 935(2):167, 2022.

\bibitem{hunter:2007}
J.~D. Hunter.
\newblock Matplotlib: A 2d graphics environment.
\newblock {\em Computing in Science \& Engineering}, 9(3):90--95, 2007.

\end{thebibliography}

\end{document}